\documentstyle{article}
\makeatletter
\newdimen\normalarrayskip              
\newdimen\minarrayskip                 
\normalarrayskip\baselineskip
\minarrayskip\jot
\newif\ifold             \oldtrue            
\def\arraymode{\ifold\relax\else\displaystyle\fi} 
\def\eqnumphantom{\phantom{(\theequation)}}     
\def\@arrayskip{\ifold\baselineskip\z@\lineskip\z@
     \else
     \baselineskip\minarrayskip\lineskip2\minarrayskip\fi}
\def\@arrayclassz{\ifcase \@lastchclass \@acolampacol \or
\@ampacol \or \or \or \@addamp \or
   \@acolampacol \or \@firstampfalse \@acol \fi
\edef\@preamble{\@preamble
  \ifcase \@chnum
     \hfil$\relax\arraymode\@sharp$\hfil
     \or $\relax\arraymode\@sharp$\hfil
     \or \hfil$\relax\arraymode\@sharp$\fi}}
\def\@array[#1]#2{\setbox\@arstrutbox=\hbox{\vrule
     height\arraystretch \ht\strutbox
     depth\arraystretch \dp\strutbox
     width\z@}\@mkpream{#2}\edef\@preamble{\halign \noexpand\@halignto
\bgroup \tabskip\z@ \@arstrut \@preamble \tabskip\z@ \cr}%
\let\@startpbox\@@startpbox \let\@endpbox\@@endpbox
  \if #1t\vtop \else \if#1b\vbox \else \vcenter \fi\fi
  \bgroup \let\par\relax
  \let\@sharp##\let\protect\relax
  \@arrayskip\@preamble}
\def\eqnarray{\stepcounter{equation}%
              \let\@currentlabel=\theequation
              \global\@eqnswtrue
              \global\@eqcnt\z@
              \tabskip\@centering
              \let\\=\@eqncr
              $$%
 \halign to \displaywidth\bgroup
    \eqnumphantom\@eqnsel\hskip\@centering
    $\displaystyle \tabskip\z@ {##}$%
    &\global\@eqcnt\@ne \hskip 2\arraycolsep
         $\displaystyle\arraymode{##}$\hfil
    &\global\@eqcnt\tw@ \hskip 2\arraycolsep
         $\displaystyle\tabskip\z@{##}$\hfil
         \tabskip\@centering
    &{##}\tabskip\z@\cr}
\makeatother

\def\beq{\begin{equation}}
\def\eeq{\end{equation}}
\def\bea{\begin{eqnarray}}
\def\eea{\end{eqnarray}}

\begin{document}

\begin{titlepage}
\begin{center}
\begin{flushright}{FIAN/TD-14/94}\end{flushright}
\begin{flushright}{December  1994}\end{flushright}
\vspace{0.1in}{\Large\bf On third Poisson structure of
KdV equation
\footnote{to appear in ''Proceedings of $V$ conference on {\it Mathematical
Physics, String Theory and Quantum Gravity}, Alushta, June 1994", Teor.
Mat.  Fiz.  1995}
}\\[.4in]
{\large  A.Gorsky
\footnote{E-mail address:
gorsky@vx.itep.ru, sasha@rhea.teorfys.uu.se}}\\
\bigskip
{\it ITEP,
Bolshaya Cheremushkinskaya 25, Moscow, 117 259, Russia} \bigskip \\
{\large A.Marshakov
\footnote{E-mail address:
mars@lpi.ac.ru,  marshakov@nbivax.nbi.dk, andrei@rhea.teorfys.uu.se}
}\\
\bigskip {\it Theory Department, P.N.Lebedev Physics Institute,
Leninsky pr. 53, 117924 Moscow, Russia\\ and\\ ITEP, Moscow, Russia}
\bigskip
\\
{\large A.Orlov
\footnote{E-mail address:
wave@glas.apc.org}
}\\
\bigskip {\it Institute of Oceanology, Moscow, Russia}
\bigskip
\\
{\large V.Rubtsov
\footnote{E-mail address: rubtsov@vx.itep.ru,
volodya@rhea.teorfys.uu.se, volodya@orphee.polytechnique.fr}}\\
\bigskip {\it ITEP, Bolshaya Cheremushkinskaya 25, Moscow, 117 259, Russia}
\bigskip
\\

\end{center}
\bigskip \bigskip

\begin{abstract}
The third Poisson structure of KdV equation in terms of canonical ''free
fields"
and reduced WZNW model is discussed. We prove that it is ''diagonalized" in the
Lagrange variables which were used before in formulation of $2d$ gravity.
We propose a quantum
path integral for KdV equation based on this representation.
\end{abstract}

\end{titlepage}

\newpage
\setcounter{footnote}0

In this letter we discuss the third Poisson structure of KdV equation
and its ''diagonalization" in appropriate variables. This is a step towards
understanding what is quantum KdV theory -- an old question  related now
to second quantized formulation of string theory, since
some particular $\tau $-functions
of integrable hierarchies of KdV type can be interpreted as effective action
in corresponding string field theory.
However, at the moment not really much is known about quantum KdV-type
systems. In particular, it is connected to the nontrivial structure of KdV (KP)
equation classical phase space.

In this short note we will propose a possible way of formulation what is
quantum KdV theory related to its {\it third} Poisson structure. As in the case
of well-known first and second
Poisson structures on KdV phase space, the third one is also related
to infinite-dimensional Lie algebras and can be realized in terms of canonical
free
fields \cite{DS,Fat-Luk,AS}. We will base a path-integral
formulation on {\it third} Poisson structure
and demonstrate that in certain (Lagrange?) variables its measure and kinetic
term acquires a very simple form of ''free fields".

\bigskip
{\bf 1.} It was found in \cite{Mag} and discussed in \cite{EOR} that in
addition to well-known first

\beq\label{1}
\{ u(x), u(y)\} _1 = \delta '(x-y)
\eeq
and second

\beq\label{2}
\{ u(x),u(y)\} _2 = \left( u(x) + u(y) \right)\delta '(x-y) + c\delta '''(x-y)
\eeq
Poisson brackets there exists a {\it non-local} third one

\beq\label{3}
\{ u(x), u(y)\} _3 = {1\over 16}\delta'''''(x-y) +{1\over 4}
\left( u(x)+u(y)\right)
\delta'''(x-y) + {1\over 8}\left( u'(x)-u'(y)\right)\delta''(x-y) +
$$
$$
+ {1\over 2}\left( u^2(x) + u^2(y)\right)
\delta '(x-y) - {1\over 4}u'(x)u'(y)\epsilon\left( x-y\right)
\eeq

We are going to demonstrate that (\ref{3}) acquires a very simple form being
rewritten in terms of variables coming from two-dimensional WZNW model
\cite{AS}

\beq\label{sl2}
g = \left(
\begin{array}{cc}
1 & F \\
0 & 1
\end{array}
\right)
\eeq
and $x(\xi ) = F^{-1}(\xi )$ \cite{P,KPZ}.

Indeed, starting from the following Poisson structure
on a space of variables (\ref{sl2}) and implying that
\footnote{Below we will use $\{ A,B\} \equiv \{ A,B\} _3$.}

\begin{equation}\label{F}
\left\{ F(x),F(y) \right\} = F'(x)F'(y)\epsilon\left( x-y \right)
\end{equation}
where

\begin{equation}
F(x) = \int_{d\xi}^{x}{\exp{\alpha\phi\left( \xi \right)}}
\end{equation}
with $ \phi\left( x \right)$ being common free field so that

\begin{equation}
u \sim \left( \phi'  \right)^2 + \beta\phi'' \equiv j^2 + \beta j'
\end{equation}
is just a Miura transform with

\begin{equation}\label{coeff}
\beta = - {2 \over \alpha}
\end{equation}
or

\begin{equation}\label{schw}
u \sim Sch F = {1 \over 2}\left\{ {F''' \over F'}  - {3 \over 2}
{F''^2 \over F'^2}\right\}
\end{equation}
for $ \phi\left( x \right)$ one has

\begin{equation}
\left\{ \phi( x ) , \phi( y )\right\} = \phi'(x)\phi'(y)\epsilon(x-y) +
{1 \over \alpha}\left\{ \phi'(y) - \phi'(x) \right\}\delta(x-y) -
{1 \over \alpha^2}\delta'(x-y)
\end{equation}
In such case for the corresponding $ U(1)$-current $ j(x) = \phi'(x)$ one
obtains

\begin{equation}
\left\{ j(x),j(y) \right\} = j'(x)j'(y)\epsilon(x-y) +
\left( j(x)j'(y) - j'(x)j(y) \right)\delta(x-y) -
$$
$$
-j(x)j(y)\delta'(x-y) + {1 \over \alpha}\left( j'(x)+j'(y) \right)\delta'(x-y)
+
{1 \over \alpha}\left( j(x)-j(y) \right)\delta''(x-y) +
{1 \over \alpha^2}\delta'''(x-y)
\end{equation}
and the Poisson bracket of Sugawara's $ u = j^2 +\beta j'$
coincides with that of (\ref{3}) provided by (\ref{coeff}).
Now, a simple remark is that if one considers the {\it inverse} function

\beq
x = f(\xi ) \equiv F^{-1}(\xi )
\eeq
then it is easy to see that for $x = f(\xi )$ and $y = f(\eta )$

\beq
\{x , y \} =  f'(\xi )f'(\eta )\{\xi ,\eta\} = {1\over F'(x)F'(y)}
\{ F(x), F(y) \} = \epsilon (x-y)
\eeq
or the bracket acquires the most simple form of canonical ''free field"
bracket for $x(\xi )$

\beq
\{ x(\xi ),x(\eta )\} = \epsilon (x(\xi ) - x(\eta ))
\eeq
or

\beq\label{can}
\{ x(\xi ), x'(\eta )\} = \delta (\xi - \eta )
\eeq
Finally, let us point out that (\ref{F}) is a sort of $r$-matrix bracket

\beq\label{r}
\{ F(x),F(y) \} = [r(x-y),F(x)\otimes F(y)]
\eeq
with

\beq\label{rr}
r_{xy} \equiv r(x-y) = \epsilon (x-y) \partial _x \otimes \partial _y
$$
$$
[r_{xy},r_{yz}] + [r_{yz},r_{zx}] + [r_{zx},r_{xy}] =
\delta _{xy}(\epsilon _{yz} + \epsilon _{zx}) + \delta _{yz}(\epsilon _{xy}
+ \epsilon _{zx}) + \delta _{zx}(\epsilon _{yz} + \epsilon _{xy}) = 0
\eeq
Eqs. (\ref{r}) and (\ref{rr}) require some comments. One should consider
(\ref{rr}) as an operator acting to {\it functions} and not to
(pseudo)differential operators. Practically it means that

\beq\label{comm}
\left[ \partial , f(x)\right] = f'(x)
\eeq
and there are no $f\partial $ terms in the r.h.s. of (\ref{comm}). In terms of
algebra $DOP(S^1)$ we are going to discuss below it means that we factor the
algebra of differential operators over ''differential" terms.

Eq.(\ref{r}) means that the variables $F$ are ``group-like'' variables and the
corresponding ``group'' is a Poisson-Lie group, i.e. the third KdV bracket is
an example of multiplicative Poisson bracket.
A similar $r$-matrix comes from the standard Lie bialgebra
structure for Lie algebra of differential operators $DOP\left( S^1\right)$
of the order $\leq 1$ extended in a proper way.
Usually a Lie bialgebra structure on ${\cal G}$ endows the dual space
${\cal G}^*$ with a Lie algebra structure in a such way that the dual map from
${\cal G}$ to $\wedge^2 {\cal G}$ is a 1-cocycle on ${\cal G}$ valued in
$\wedge^2 {\cal G}$.
As a coalgebra for the Lie algebra ${\cal G} = DOP (S^1)$ we have to take the
linear space
generated by quadratic and linear differentials with a ``cocentral''element
$\log\partial $.
The elements of the coalgebra should be written in the form

\beq\label{coal}
K\log\partial + \partial^{-1}a + \partial^{-2}b
\eeq
and all terms with higher negative degrees should be truncated in all
commutators.
The commutation relations in the $DOP\left(S^1\right)$  are the following

\beq\label{dop}
[f\partial +g,h\partial +e] = \left(fh' - hf'\right)\partial + fe' -hg'
 +  {\hat C}c\left(f\partial +g,h\partial +e\right)
\eeq
where(cf \cite{AdKKP})

\beq\label{cocycl}
c\left( f\partial +g,h\partial +e\right) = {1\over 6}Res\left( fh'''\right) -
{1\over 2}Res\left( fe'' - hg''\right) - Res\left( ge'\right)
\eeq
Duality is given by the standard Adler-Manin residual trace.

\beq\label{dual}
\langle f\partial + g + c{\hat C}, K\log\partial + \partial ^{-1}a
+ \partial ^{-2}b \rangle = cK + Res(ga) +Res(fb)
\eeq
The only non-trivial commutator in the coalgebra is

\beq\label{com}
[\partial^{-1}a,\log\partial ] = -\partial^{-2}a'
\eeq
Therefore the co-bracket $\delta : DOP\left(S^1\right) \to
\bigwedge^2\left(DOP\left(S^1\right)\right)$
is following

\beq\label{cobra}
\delta({\hat C}) = \delta\left(g\right) = 0
$$
$$
\delta\left(f\partial\right) = f'\otimes {\hat C} - {\hat C}\otimes f'
\eeq
Hence,

\beq\label{comp}
\langle\delta\left(f\partial\right) , \partial^{-1} a\otimes\log\partial
\rangle =
\langle f\partial , [\partial^{-1} a, \log\partial]
\rangle =
- \langle f\partial ,\partial^{-2}a' \rangle  = - Res\left(fa'\right)
\eeq
The co-bracket (\ref{cobra}) can be written with the help of $r$-matrix
$\delta (x) = [r,\Delta (x)]$ where $x \in {\cal G}$ and
$r \in {\cal G}\otimes {\cal G}$. Indeed, if one takes
$r_{xy} = 2\epsilon (x-y)$, then

\beq\label{r?}
[\Delta (f\partial ), {r}] = 2[f(x)\partial _x + f(y)\partial _y,
\epsilon (x-y)] =
$$
$$
= 2\left( f(x) - f(y)\right)\delta (x-y) -
{\hat C}Res(f(x)\delta '(x-y)) +
{\hat C}Res(f(y)\delta '(y-x)) =
$$
$$
= {\hat C}\left( f'(x) - f'(y)\right) = f'\otimes {\hat C}
- {\hat C}\otimes f'
\eeq
Another remark links the expression (\ref{F}) with integrable systems
is that for the bracket (\ref{3}) the functional $\int{u(x)}$ is
a Hamiltonian for the hamiltonian form of KdV equation.
Using (\ref{schw}), we have that

\beq\label{sch}
H\left(F\right) = \int dx Sch\left( F\right)
\eeq
is a Hamiltonian for Ur-KdV in the coordinates of \cite{AS}.
The equation is

\beq\label{urk}
F_t = F''' - {3\over 2}{\left(F''\right)^2\over F'}
\eeq
and can be rewritten in Hamiltonian form using the Hamiltonian (\ref{sch}) and
(non-local) operator

\beq\label{o-1}
\Omega ^{-1} = -{1\over 2}{F'\partial^{-1}F'}
\eeq
which is equivalent to the usual KdV written in the hamiltonian form with
respect of the third structure (\ref{3}). These structures were discussed in
\cite{Wilson}.

The ``group'' nature of the variables would imply some natural description of
the commuting hamiltonians for this equation in terms of central elements of
the correspondig coboundary Poisson-Lie ``group''.

\bigskip
{\bf 2.} Up to now we have shown that the third Poisson structure is
diagonalized in terms of $x(\xi ) = F^{-1}(\xi )$ variables having clear sense
of one-dimensional or two-dimensional {\it holomorphic} reparameterizations.
These variables are called in hydrodynamics as the Lagrange variables in
contrast to the Euler variables $u(x)$, corresponding to the field of
velocities of particles of a liquid in a fixed point. On the contrary, the
Lagrange variables are nothing but co-ordinates of a fixed ''particle" being
exactly $x(t,x_0=F)$, i.e. coincide with the Polyakov variables \cite{P,KPZ}.
Remarkable enough the same variables arise and make sense in two
{\it a priori} physically different problems.

To formulate the path integral for KdV equation the only additional remark we
should make is that the natural measure is also written in terms of Lagrange
variables. Indeed, if one takes for example the measure
coming from WZNW model \cite{AS}, then

\beq\label{measure}
\prod {dF(x) \over F'(x)} = \prod dx(\xi )
\eeq
and the path integral looks as follows

\beq
Z = \int Dx \exp i\left( S_0 + H \right)
\eeq
where

\beq\label{into}
S_0 = \int _{dt}\theta = \int _{dt} \delta ^{-1}\Omega
$$
$$
\Omega = \int _{dx\wedge dy} {\delta F(x)\over F'(x)}\wedge
{\delta F(y)\over F'(y)}\delta '(x-y) = \int _{dx} {\delta F(x)\over
F'(x)} {d\over dx}{\delta F(x)\over F'(x)}
$$
$$
=  \int _{d\xi} \delta x(\xi ){d\over d\xi}\delta x(\xi )
\eeq
$H$ is a Hamiltonian

\beq
H(x) = \int _{dtd\xi} x'Sch F = -{1\over 2}\int _{dtd\xi}{x''(\xi )^2\over
x'(\xi )^3}
\eeq
and $Dx = \prod dx$ is {\it ''free" measure} (\ref{measure}).
Finally, integrating $\delta \theta = \Omega $ one gets

\beq
\theta = {1\over 2}\int _{d\xi}x'(\xi )\delta x(\xi )
\eeq
so that

\beq\label{freea}
S_0 = \int _{dtd\xi}\theta = {1\over 2}\int _{dtd\xi} \dot{x} x'
\eeq
It is easy to check the consistency of (\ref{into}) and (\ref{freea}) with
the formulas (\ref{F}), (\ref{can}) and (\ref{o-1}).

\bigskip
{\bf 3.} In this letter we have proved that naively non-local third Poisson
structure
of the KdV equation acquires a simple and physically clear form if one works
in appropriate variables. The choice of convenient variables allows us to write
formal path integral in the Hamiltonian form which can be treated as a possible
starting point to study quantum KdV theory.

In spite of growing interest to this problem we should mention that it is
still an open question. One of the problems is unclear at the moment physical
interpretation of the possible result or in other words what should we
expect from correlators in quantum KdV theory?

The existence of many different Hamiltonian formulations of classical KdV
equation imply a natural question of relations between the path integrals
based on different classical formulations. In particular, it is interesting to
compare the path integral proposed above with the more common approach to
KdV equation relation to coadjoint orbit quantization of the Virasoro algebra
resulting in another path integral \cite{GOS}.

Finally, let us make a remark concerning the ''full" path integral. In fact,
the exact relation between Euler and Lagrange variables looks as

\beq\label{schb}
u(x) = b[F(x)]F'(x)^2 + Sch(F)
\eeq
and includes ''initial data" $b$ which corresponds to the choice of different
orbit \cite{AS}. It seems natural to think that the full path integral for KdV
theory should include also integration over different orbits, i.e.

\beq\label{fullint}
Z = \int DbDx \exp i\left[ S_0(x) + H(b,x)\right]
\eeq
with some (unknown) measure $Db$.
We are going to return to these questions elsewhere.

\medskip
We are grateful to B.Enriquez, D.Gurevich, S.Kharchev, A.Niemi, K.Palo and
A.Rosly for illuminating discussions. The work was partially supported by
ISF grant MGK000 and grants of Russian fund of fundamental research
RFFI 93-02-03379 and RFFI 93-02-14365.
The work of A.M. was partially supported by NFR-grant No F-GF 06821-305 of the
Swedish Natural Science Research Council.
V.R. is grateful to Centre Mathematiques of Ecole Polytechnique (Palaiseau)
for warm hospitality and to CNRS for a financial support in the last stage of
this work.


\begin{thebibliography}{10}

\bibitem{Mag}
F.Magri {\cal Journ.Math.Phys.}, {\bf 19} (1978) 1156-1162

\bibitem{EOR}
B.Enriques, A.Orlov, V.Rubtsov, JETP Lett., October 1993

\bibitem{AS}
A.Alekseev, S.Shatashvili  Nucl.Phys. {\bf B329} (1989) 719

\bibitem{P}
A.Polyakov  {\cal Mod.Phys.Lett.} {\bf A2} (1987) 893

\bibitem{KPZ}
V.Knizhnik, A.Polyakov, A.Zamolodchikov
 {\cal Mod.Phys.Lett.} {\bf A3} (1988) 819

\bibitem{DS}
V.Drinfeld, V.Sokolov, {\cal Journ.Sov.Math.}, {\bf 30} (1985) 1975-2036

\bibitem{Fat-Luk}
V.Fateev, S.Lukyanov, Int.Journ.Mod.Phys. {\bf A3} (1988) 507;
{\bf A7} (1992)

\bibitem{Lukyanov}
S.Lukyanov, {\cal Funct.Anal.appl.} {\bf 22} (1988) 1

\bibitem{Wilson}
G.Wilson, {\cal Phys.Lett.} {\bf A132} (1988) 445

\bibitem{AdKKP}
E.Arbarello, C.de Koncini, V.Kac, G.Procesi, {\cal Comm.Math.Phys.} {\bf 117}
(1988) 1

\bibitem{GOS}
A.Gorsky, {\cal Yad.Fiz.} (1991)


\end{thebibliography}
\end{document}